# Towards an Interoperable Identity Management Framework: a Comparative Study

Samia El Haddouti[1] and Mohamed Dafir Ech-Cherif El Kettani[2]

[1] Information Security Research Team – ISeRT, ENSIAS, Mohamed V University
Rabat, Morocco

[2] Information Security Research Team – ISeRT, ENSIAS, Mohamed V University
Rabat, Morocco

**Abstract**

The development of services and the growing demand for resources sharing among users from different organizations with some level of affinity have motivated the creation of Identity Management Systems. Identity Management has gained significant attention in recent years in the form of several projects producing many standards, prototypes and application models both in the academia and the industry. However, the interoperability between different Identity Management Solutions is still a complex challenge yet to achieve. The user can only use one Identity Provider within a single Service Provider session, when in many scenarios the user needs to provide attributes from multiple Identity Providers.

This paper presents the state of the art of our researches and it focuses on two main topics: first, to provide a detailed study about the Identity Management and the integrated disciplines and technologies in general; secondly, to summarize the main approaches that have been proposed to overcome the interoperability challenge.

## 1. Introduction

Identity management and the integrated technologies play a big role to recommence administrative processes and promote e-government development by bringing services closer to citizens. People use the internet to manage finances, to access resources, for shopping, to communicate and so on. Each activity involves interacting with a Service Provider. Only users with proper privileges can access the controlled services and resources. In other terms, a registration phase is essential to allow users to receive credentials that are required in case they want to access those services.

To check the user's privileges, the application providing the service must verify the user's identity. Prior to the introduction of Identity Management systems, Service Providers handle this by themselves. However, this practice has many drawbacks. On the one hand, the number of passwords or tokens increases in a linear fashion to the number of Service Providers which is not user-friendly (user need to remember many passwords), and besides, it is not efficient for the business as they cannot tell whether the same customer uses multiple services. On the other hand, to grant the access to services, a Service Provider requests and stores personal attributes. However, many Service Providers are not sure about the correctness of attributes that are disclosed by the user during registration.

Several technologies and frameworks have been developed to carry out the necessary activities related to the Identity Management and to provide mechanisms by which the end users can manage their identities. The implementation of the common processes across multiple accounts will standardize and simplify procedures, and will reduce mistakes and cost. In addition to the technological decisions, Identity Management also focuses on security and privacy aspects to protect consumers by ensuring the authenticity and reliability of the provided information.

Unfortunately, one of the limitations of the current Identity Management Systems is that the user can only select one of his Identity Providers in any given session with a Service Provider. For many web based services this is not enough. Users need to select attributes from multiple Identity Providers in a single service session. Furthermore, end users may have multiple accounts on different social or access networks (e.g. Google, Facebook, Twitter, etc...). For each one of these accounts, the user holds credentials to perform the authentication process and some attributes describing the information the services need to know about the user. However, for the same user, a lot of attributes can be replicated on different Identity Providers making difficult to manage these attributes in consistent way. Several initiatives put its research efforts into building a more consistent view of Identity Management taking into account the interoperability requirement.

This work is structured as follows. Section 2 defines the basic concepts of Identity Management. Section 3 presents the evolution of Identity Management Systems. Section 4 introduces current approaches related to Identity Management existing frameworks. Section 5 presents the requirements of an Identity Management System and a comparative study of the four currently popular User-Centric Identity Management systems. Section 6 treats the interoperability challenge of current Identity Management Systems and it provides an overview of the proposed approaches for an advanced Identity Management. Section 7 serves as conclusions and future work.

## 2. Basic Concepts of Identity Management

The Internet has provided a great flexibility in the interactions between Service Providers and their users. Authorization mechanisms ensure that only authorized users can gain access to the protected resources. This suggests the need for a form of digital identities for users and a way to manage these identities.
Identity Management is a set of functions and skills such as the management, the discovery and the exchange of information which is used to facilitate the establishment of security mechanisms. The authentication which is an integral part of Identity Management, serves to verify claims about holding specific identities. Identity Management is therefore fundamental and it includes other security constructs such as authorization and access control [1].

2.1 Components of an Identity Management System

An Identity Management System includes the following components [2]:
- **Subject:** is a party, typically individuals, who wants to access a service.
- **Digital Identity:** There are several different definitions of the identity in the context of Identity Management. Pfitzmann and Hansen [3] define the identity as: "An identity of an individual person may comprise many partial identities of which each represents the person in a specific context or role. A partial identity is a subset of attribute values of a complete identity, where a complete identity is the union of all attributes values of all identities of this person".
- **Identity Provider (IdP):** is the entity responsible for managing and issuing identities for users in order to interact with an Identity Management system. The IdP can also provide additional information (attributes) about the user to the resource upon resource's request.
- **Service Provider (SP):** is the entity that provides resources and services to the user. This last one has to be authenticated first in order to access the content of resources. An access control decision is made by the SP based on the retrieved information about the user.

## 3. Evolution of Identity Management Systems

Identity management systems models are classified as conventional, centralized, federated and user-centered [4].

3.1 Conventional Model

This model requires that each user possess an identifier to access each isolated service. For this model, a SP plays also the role of an IdP. This approach tends to be costly for both users and SPs. Each SP requires its own attributes to form the user's identity so that the user has to provide the same information as many times as the number of accounts created in the SP.
From the security's point of view, with multiple accounts and so many passwords, it's becoming increasingly difficult to remember them all by the user so that the same password may be registered in multiple providers giving a rise to security risk from identity fraud and other forms of criminal activity.

3.2 Centralized Model

In order to avoid the redundancies and inconsistencies in the conventional model, the centralized model was appeared as a possible solution based on the sharing of user identities among SPs and on the concept of single authentication (SSO: Single Sign-on). In this approach, a central IdP became responsible for the collection and the provisioning of the user's identity information so that the user has to remember only a single set of credentials to access different services. Drawbacks of the centralized model derive from the fact that it has a single IdP which represents a single point of failure and a central point with full control over user data. The user has no control anymore on which data are stored or actually transmitted to the SP.

### 3.3 Federated Identity Model

The federated identity model was introduced to separate account management from the service itself and to decentralize the responsibility of one IdP to multiple such IdPs which have a trust relationship amongst each other. These IdPs are arranged in different administrative domains. An administrative domain can represent a company, a university, and so on. Each administrative domain is composed of users, multiple SPs and a single IdP. Typically the trust between IdPs and SPs is made explicit by signing policies and agreements that describe the requirements and responsibilities of both the IdPs and SPs. The agreements between providers ensure that identities issued in a domain will be recognized by SPs in other domains. In this way, the concept of single sign-on is provided even when different domains are involved with user accesses.

Typical examples of Identity Federations are the federations as operated by National Research and Education Networks (NREN) such as IDEM www.idem.garr.it by the Italian NREN - GARR, AAF aaf.edu.au by the Australian NREN - AARNET, and eduIDM www.eduidm.ma by the Moroccan NREN-MARWAN.

In most of federated identity systems, the user only has limited (or even no) control about the attributes that are exchanged between the IdP and the SP.

### 3.4 Users-Centric Identity Model

Currently, emerging paradigm of Identity Management is user-centric identity model. David Recordon VeriSign Inc and Drummond Reed give this definition of user centricity: "User-centric Identity Management is understood to mean digital identity infrastructure where an individual end-user has substantially independent control over the dissemination and use of their identifier(s) and personally-identifiable information (PII)."[5].Another definition is given by Tewfiq El Maliki and Jean-Marc Seigneur: "In user-centric identity management the user has the full control over his/hers identity and consistent user experience during all transaction when accessing his/her services." [6].

In User-Centric identity model, the user herself always remains the owner of her identity data. Identity data are managed and stored within the user's domain, usually on a secure token such as a smart card and are transferred to the SP only if the user explicitly gives her consent to do so. Many countries use the electoric Identity (eID) technology with smart (or SIM) cards. However, only static attributes (i.e. personal properties that do not change during a user's lifetime such as name, date of birth, etc...) can be stored on these cards. Moreover, users often have little impact on the attributes that are released during authentication. In some architecture, they are always identifiable and need to release attributes that are not required for the particular service.

## 4. Selected Identity Management Systems

Several frameworks for Identity Management exist; each of them has its own distinguishing features.

### 4.1 Shibboleth

"Shibboleth is a standard based open source software package for web single sign-on across or within organizational boundaries. It allows sites to make informed authorization decisions for individual access of protected online resources in a privacy-preserving manner."[7]. The Shibboleth project was an initiative of the Internet2 consortium in 2000 and was quickly adopted by research and education communities. Version 1.0 was released in 2003 and on 1 July 2015, the Shibboleth project has announced the release of V3.1.2 of the IdP software. The Shibboleth software implements widely used federated identity standards, principally the OASIS Security Assertion Markup Language (SAML), to provide a federated single sign-on and attribute exchange framework. There are three main roles within Shibboleth software: IdP, SP and Discovery Service (DS) also called "Where Are You From" (WAYF).

### 4.2 Liberty Alliance

The Liberty Alliance project has emerged as a consortium of companies from different areas such as telecommunications, banks, universities, etc…, in order to establish standards, specifications and best practices for Identity Management in computer systems. These specifications were addressed to the integration with Web services applications [8], [9].

The proposed framework for identity management is composed of three main components: Identity Federation Framework (ID-FF), Identity Web Services Framework (ID-WSF) and Identity Services Interface Specifications (ID-SIS).

### 4.3 OpenID

OpenID is an open and decentralized standard for Identity Management [10], [11]. The basic idea of OpenID is that users create accounts by selecting an OpenID IdP, and then use these accounts to access any service that accepts OpenID authentication.

OpenID authentication provides a way to prove that an end user controls an identifier. This identifier (or handle) is usually a URL (Uniform Resource Locator) or, in some cases, an XRI (Extensible Resource Identifier).

As SAML 2.0 webSSO Profile, OpenID authentication is based on SSO and uses only the standard HTTP(S) to transmit authentication results between IdPs and SPs. it does not require any special capabilities of the User-Agent or other client software.

OpenID authentication specification and the SAML Web browser profile appear to offer very similar functionalities. However, there are differences between SAML and OpenID at the discovery mechanism of IdPs and the expressiveness of data generated and processed in identity transactions. On the other hand, SAML is based on an explicit trust between SPs and IdPs which is not the case for openID. A detailed comparison of the OpenID and SAML is illustrated at [12].

Nowadays, many organizations and SPs are using OpenID authentication to provide the access to their services. Google, LiveJournal, Facebook, Yahoo, Microsoft, AOL, MySpace, Sun, IBM, PayPal, are examples of these organizations and SPs.

### 4.4 CardSpace (InfoCard)

The system CardSpace, originally called InfoCard, is a platform component of Microsoft designed to offer users a consistent support for handling with multiple digital identities by adopting the federated user-centric identity meta-system [13]. This approach provides a consistent way to work with multiple digital identities using any type of security token, including simple usernames tokens, X.509 certificates, Kerberos tickets, SAML tokens, or any other token. It is a technology that helps developers to integrate consistent identity infrastructure into applications, Web sites, and Web services.

## 5. Requirements of an Identity Management System

Before the adoption of an Identity Management System, a set of requirements need to be taken into account by organizations to assess which system should be deployed.

After the failure of Microsoft's Passport system, Kim Cameron discussed the issues and thought about what is needed to build a successful Identity Management System. One of the results of his researches was his seven laws of identity [14]. Furthermore, we will add a set of requirements addressing functional and business concerns.

### 5.1 User Control and Consent

When a SP requests an IdP to release a personal information about an end user, this last one should approve whether such information could be released or not. Thus, the system must be designed to only reveal identity data with the user's control and consent.

### 5.2 Minimal Disclosure for a Constrained Use: Data Minimization

The Identity Management System should be built to disclose no more than the necessary identifying information. Thus, only the minimum amount of personal data is stored. A system built with this feature is therefore a less attractive target for identity theft.

### 5.3 Justifiable Parties

The Identity Management System must make its user aware of the party or parties with whom she is interacting while sharing information. The disclosure of identifying information is limited to parties having a necessary and justifiable place in a given identity relationship. Only those parties authorized to access the data.

### 5.4 Directed Identity

The main idea is that the system must be capable of supporting a range of identifiers with varying degrees of observability and privacy. Users do not want everyone to know their identifiers. They prefer to keep them private. However, public web sites and commercial organization want everyone to know their identifiers and hence be able to contact them. Therefore, users must be able to use the omnidirectional identifiers provided by public entities in order to confirm who they are dealing with and to ensure that their personal information is being disclosed appropriately. At the same time, unidirectional identifiers (private identifier) should be assigned for use in a specific communication in order to minimize data linkage across different sites.

### 5.5 Pluralism of Operators and Technologies

It will be crucial for any Identity Management System to have a good degree of compatibility with other existing systems to make it a hugely successful one. Users' identities should be represented in a common format in order to allow an easy understanding and validation of them even in the face of multiple administrative domains. Hence, the Identity Management Systems have to support the extensible mapping between identities.

### 5.6 Human Integration

Securing the link between the user and a machine is an essential component to offer strong protection against identity attacks. Moreover, the user should understand the *ceremony* of all communications as described by Carl Ellison [15].

### 5.7 Consistent Experience Across Contexts

This law enables users to have a consistent experience when they are switching between technologies. In this way, users are able to manage their identities with a transparency.

### 5.8 Security

An Identity Management System must provide a sufficient level of security of its services against attacks [16]. The basic security mechanisms are:

- *Authentication:* is used to ensure the identity of a user or a device for the purpose of controlling the access to services. This process can be done either by a user-id and password for simple web services or by OTP (One-time password) and hardware tokens for more secure services.
- *Integrity*: refers to protecting data from being modified by unauthorized parties during processing or transmission.
- *Confidentiality:* refers to limiting information access and disclosure data to authorized parties. The encryption is the key component to protect confidentiality of information.
- *Non-repudiation:* is the ability to prove that if a transaction has taken place, the sender of a message cannot later deny the sending of a message and the recipient cannot deny the reception of this message.

### 5.9 Privacy

As an Identity Management System will manage personal data of individuals and other data that is applied for authentication or authorization, the maintenance of privacy is vitally important. Identity Management System can be built according to the Privacy Enhancing Technologies (PETs) [17] to guarantee the privacy for a system. Some requirements that can be used to ensure the privacy of a user in the Identity Management setting [18] are:

- *Use of Pseudonyms*: The use of pseudonyms as identifiers helps to have the anonymity in Identity Management. The user is not known to a service-provider by his identity (name, address, city), but by series of characters (letters, numbers and punctuation marks). Thus, Relaying Parties cannot exchange information about individual users. An essential factor for effectiveness of pseudonyms is the unlinkability between the pseudonym and its holders.
- *Anonymity:* is defined in terms of the linkability of items of interest that are any distinct features that might reveal information about users. Examples of items of interest include names, e-mail messages, and search engine queries. Furthermore, the user's identity and real name may themselves be considered items of interest. Thus, the information provided by the user to set a digital identity should not be used to discover any other of his identities. The use of pseudonyms is a way to ensure anonymity. The pseudonym should be unlinkable to the original partial identity and the system should offer the possibility of creating, updating and deleting different pseudonyms.
- *User-controlled linkability:* is the core concept of the Identity Management. It aims to realise unlinkability of different user's actions. Thus, the communication partners involved in different actions of the same user cannot aggregate the personal data disseminated during these actions.
- *User Empowerment:* this feature allows users to discover his privacy rights. Hence, users should be empowered to control how much of their identities to share, under what conditions and for what purpose.
- *Remote Administration of User Policies*: To enforce the control and being aware of personal data released to other parties, The Identity Management System should allow users to administrate their data remotely.
- *Usage of Privacy Standard:* To gain more control over the use of personal information on Web sites visited by users, the privacy standard enables Web sites to express their privacy practices to their visitors in a machine readable format. On 28 January 2002, the W3C released a proposed specification of the Platform for Privacy Preferences Project (P3P) as an industry standard.

### 5.10 Trustworthiness

As the user will provide personal information to an Identity Management System, it is a prerequisite that the user builds trust relationships with this system [16]. The main factors to maintain a mutual trust between a user and Identity Management System are:

- *Trusted Seals of Approval:* security and privacy seals can be used in an attempt to reassure the user that the system is going to handle user data in the agreed way and according to standards. Thus, an Identity Management System can be considered truly secure.
- *Using Open Source Technologies:* being able to review and audit the source code of a system to understand and validate its security and privacy properties, provides additional ways to evaluate the trustworthiness of this system.

- *Segregation of power:* a separation of the power is one of the basic functionality to gain the trust. With this property, an entity will not have a dominant position over other entities so that it cannot abuse its power to monopolize a service and users can choose a supplier based on their performance.
- *Legal Protection:* users need to feel more comfortable to get involved in transactions especially when it comes to financial transactions such us e-banking, web-commerce, e-taxation, etc... Legal Protection is another way to achieve user-trust which in turn increases the trustworthiness towards the system.

### 5.11 Usability

Usability, as defined by ISO Standards for Usability: "The extent to which a product can be used by specific users to achieve specified goals with effectiveness, efficiency and satisfaction in a specified context of use." [19].
As Identity Management System deals with the management of end-user identities, and therefore requires frequent interactions with end users, usability is the first crucial challenge to be addressed. Usability is particularly crucial in recent user-centric solutions whose underlying design principle is that users must be in control of their identity information. It is well known that poor usability implies a weakness in authentication.

### 5.12 Identity Recovery

An Identity Management System should specify an identity recovery mechanism to recover a digital identity if it has been stolen by an intruder [16].

### 5.13 Context Detection

This functionality describes possibilities to detect more information about the user's environment to classify situations in order to determine which personal data should be disclosed and to make suggestions for further activities according to the current situation of the user [20].

### 5.14 Location Independence

This requirement deal with mobility and it allows a remote access to the Identity Management Systems from different locations without any restrictions [16].

### 5.15 Identity Administration: Creating, updating and deleting Identity and its related information

The System should provide users with mechanisms to create update and delete her existing partial identity. [16]

### 5.16 Digital Evidence

Digital evidence is a mechanism that uses a digital data as witnessing source to claim liability or legal protection in case of identity theft, wrong delivery, unauthorized access, and so on. [21]

### 5.17 Data Retention

Data retention policies should be established and implemented to retain persistent data securely as long as needed. [22]

### 5.18 Affordability

Affordability is another factor allowing a wide-spread adoption of a system. The integration of an Identity Management System should not be more expensive than the actual transactions. Furthermore, it might be advantageous if a new system could bring additional advantages by creating the possibility for new business models and services [16]. Among the requirements that would be helpful for any a new system to get more adoption, we find:

- *Flexible Business Model:* an Identity Management System supporting several deployments is essential for people's daily lives. As a system usually interacts with business organizations, it will be important to offer a substantial amount of incentives to get more adoption.
- *Powers of market:* an Identity Management System providing a diversity of services with the ease of availing these services is able to reach a remarkable penetration of market.
- *Subsidies for development, use, operation, etc:* in case the Identity Management System is in line with the governmental objectives, it can benefit from the subsidies for development, use, operation etc...

### 5.19 Reducing System's Complexity

An Identity Management System should be developed with the simplicity concept. Even if a system may be very complex in its architectures, it is wise to hide this complexity from the user with a simple and intuitive User-Interface. By reducing the complexity of the system, the usability will be increased.

A Comparative Analysis

The total set of requirements presented above is used as comparable metrics for comparing the four currently popular user-centric Identity Management Systems: OpenID 2.0, Shibboleth, Liberty Alliance, and CardSpace. Our findings, based on respective specifications, documentations, wiki pages, webpages and published papers, are presented in the table below.

We have used the tick "√" to indicate that the system satisfies a respective requirement and the character "X" to indicate that the system does not satisfy the requirement. The dash "—" character has been used in cases where there is no way to quantify this requirement.

Table 1: A comparative analysis of User-Centric Identity Management Systems

| | | CardSpace | Liberty Alliance | Shibboleth | OpenID |
|---|---|---|---|---|---|
| User Control and Consent | | √ | X | X | √ |
| Data Minimization | | √ | √ | √ | √ |
| Justifiable Parties | | √ | √ | √ | √ |
| Directed Identity | | √ | √ | √ | √ |
| Pluralism of Operators and Technologies | | √ | X | X | X |
| Human Integration | | √ | X | X | X |
| Consistent Experience Across Contexts | | √ | X | X | X |
| Security | Authentication | √ | √ | √ | √ |
| | Integrity | √ | √ | √ | √ |
| | Confidentiality | √ | √ | √ | √ |
| | Non-Repudiation | √ | √ | √ | √ |
| Privacy | Use of Pseudonyms | √ | √ | √ | X |
| | Anonymity | X | X | X | X |
| | User Controlled Linkability | X | X | X | X |
| | User Empowerment | √ | √ | √ | √ |
| | Remote Administration of user Policies | X | X | X | X |
| | Usage of Privacy Standard | X | X | X | X |
| Trustworthiness | Trusted Seals of Approval | X | X | X | X |
| | Using Open Source Technologies | X | -- | √ | √ |
| | Segregation of power | X | X | X | √ |
| | Legal Protection | -- | -- | -- | -- |
| Usability | | √ | X | X | X |
| Identity Recovery | | √ | X | X | X |
| Context Detection | | √ | X | X | X |
| Location Independence | | X | √ | √ | √ |
| Identity Administration | | √ | X | X | √ |
| Digital Evidence | | X | X | X | X |
| Data Retention | | X | X | X | X |
| Reducing System's Complexity | | -- | -- | -- | -- |
| Affordability | Flexible Business Model | -- | -- | -- | -- |
| | Powers of market | √ | -- | √ | √ |
| | Subsidies for development, use, operation etc. | -- | -- | -- | -- |

Interpretations and Synthesis

- Each Identity Management solution has its benefits and downsides. As evident from the table above, CardSpace has met the maximum number of requirements and is followed by Shibboleth, OpenID and then Liberty Alliance.
- The existing Identity Management solutions have somewhat good support for security. However, many of them fail substantially to meet many privacy requirements especially Anonymity, User-controlled Linkability, Remote Administration of user Policies and Usage of Privacy Standard.
- As is illustrated on the table above, the existing Identity Management solutions don't take into consideration the Context Detection requirement. Namely that this requirement play a crucial role in Mobile Identity Management.
- The usability requirement, that means the ease of deployment and of use, is not insured by the current Identity Management Systems (except for CardSpace).
- Human Integration requirement that provides a strong protection against identity attacks by securing the link between the user and the machine is missed in current solutions (except for CardSpace).
- Digital Evidence and Data Retention ensure the liability of the system. However, these functionalities are missed in current Identity Management Systems.
- Requirements: Legal Protection, Reducing System's Complexity, Flexible Business Model and Subsidies for development, use operation etc, are important for widespread usage and reputation for an Identity Management System. However, there are no ways to quantify these requirements for current systems.

## 6. Interoperability challenge of Identity management Systems

6.1 Open Issues

The interoperability of Identity Management Systems is one of growing concern. When using identities as a means of controlling access to ever-larger online and public information systems, especially e-government and e-business systems, the issue of interoperability becomes a crucial one.

Nowadays, end users have multiple accounts on different social and access networks (e.g. Google, Facebook, the local internet service provider, etc...). For each one of these accounts, a user holds credentials to perform the authentication process, and some attributes describing the information the services know about the user. However, for the same user, a lot of attributes can be replicated on different IdPs making difficult to manage these attributes in consistent way.

On the other hand, some particular service interactions require the use of identity information coming from different sources (different IdPs). For example, a contact from one account, age from other and address from a third one. This implies the need to communicate data across different domains by using different identity tokens, protocols, standards, and so forth. However, the current

Identity Management Systems have one significant limitation. The user can only use one IdP within a single SP session while in many web based services the user needs to select attributes from multiple IdPs.

## 6.2 Related Work

Several approaches have been proposed to overcome the interoperability challenge.

In [22], authors present a description of the SWIFT Identity Management framework. The SWIFT project leverages Identity Management as a key technology of the future internet, tackling problems like the integration of the network and application layer. This framework describes how identity aggregation, cross-layer and pseudonymity features can be addressed to provide the end user with the required mechanisms to use his identity information to access any service, no matter if it is a web service or the network access service. As IdPs can support different functions depending on the services they provide, three roles of IdPs are distinguished: authentication providers, attribute providers, and identity aggregators. Moreover, the framework allows users to use Virtual Identities (VID) which is a special kind of digital identity built up as the aggregation of attributes and credentials from different sources (providers) allows overcoming the interoperability issue of current Identity Management Systems. However, there are issues that could be improved in this approach. The paper [22] does not provide detailed security and privacy analysis and the framework does not resolve the problem of naming heterogeneity that occurs when combining sets of attributes. Moreover, the identity aggregator could be seen as a central point of failure.

The Linking Service (LS) is a special kind of aggregation entity proposed in [23]. The LS is a new web service acts as an intermediary between the IdP and SP creating links through user interaction in order to achieve attribute aggregation. Each IdP knows one partial identity of the user and no IdP is aware of any of the other user's partial identities. On the other hand, the LS only knows that a user is known to several IdPs, and it holds the links to these on behalf of the user without knowing who the user is. Privacy preservation is ensured through a minimal of trust. The user, IdPs and SPs trust the linking service to hold the links securely and to only divulge them to SPs under the instructions of the user.
This conceptual model satisfies most of requirements of an Identity Management Systems and is beneficial with regards to obtaining data from various sources. However, this approach is based on web services and does not deal with network services or cross-layer support. Furthermore, it does not provide a detailed security analysis and it does not offer the possibility to hide the original source of data since assertions are signed by source IdP instead of by the LS. In the context of the clients' identity verification process, this approach does not resolve the problem of naming heterogeneity that occurs when combining sets of attributes.

The work proposed by [24] enriches the SWIFT project represented by [22] with privacy and security missing part. The paper describes an advanced management infrastructure able to provide end user with pseudonymity, identity aggregation, cross-layer SSO and advanced authorization decisions. The security analysis has been performed with widely used tool AVISPA. The results of this analysis demonstrate that the proposed framework fulfills the requirements of privacy, pseudonymity and unlinkability. Nevertheless, the problem of naming heterogeneity still persists.

In [25], authors present a new approach for user-centric Identity Management using trusted modules. This model is based on a trusted secure element which acts as a gateway between IdPs and SPs. The proposed approach enriches the LS concept and it tackles several privacy and security problems of current Federated Identity Management Systems and current electonic Identity (eID) technology initiatives. On the one hand, an IdP cannot profile the user's actions, as there is no direct link between IdPs and SPs. On the other hand, the disclosure of personal information is controlled by multiple parties, preventing any single entity from compromising user privacy. In addition, explicit user consent is required prior to the release of data and users can restrict the disclosure of personal information. Each user can configure its own privacy policy. Despite the benefits of this proposed solution, there are issues that could be improved in this approach. Firstly, it does not take into account the problem of naming heterogeneity. Secondly, it does not provide a global Identity Management System covering from network layer to high level services. It is only focused on web services.

In [26], authors propose a solution consisting of obtaining strong identifiers by combining user attributes within IdPs using direct attribute matching and ontologies in order to find correspondences in users' attributes distributed on IdPs, and to solve Schema-Level conflicts arise when similar concepts are labelled in a different way, or when different concepts are labelled in a similar way. The proposed approach is based on a mechanism named User Identification Strengthen (UsIdS) that performs an open search through users IdPs finding correspondences in the users' attributes. The privacy is taken into account for this approach by defining a protocol for the communication process between UsIdS and IdPs in order to assure that user attribute values are not disclosed when IdPs network establishment is being perform. However, the security aspect is not treated. Furthermore, the proposed approach does not deal with network services or cross-layer support.

## 7. Conclusion and Future Work

Along this paper we have we have reviewed most important concepts underlying the Identity Management. Each Identity Management solution has its benefits and downsides. Despite the diversity of prototypes and application models developed to carry out the administration and the management of identities, users can only select one of their IdPs in any given session with a SP. For many scenarios, this is not enough and the user needs to select or aggregate the attributes from several IdPs in order to justify his authorization to access the requested resource.

As mentioned earlier, approaches used for attribute aggregation are beneficial with regards to obtaining data from various sources. However, there are issues that could be improved in these approaches. In spite of findings related to defining ways of providing security and privacy properties, naming heterogeneity and cross-layer SSO, a solution that integrates all these properties has yet to be found. Our future work will cover the missing parts of the most recent proposals on Identity Management towards a unified model which will overcome the interoperability challenge by providing the aggregation of attributes from multiple sources, without a necessity to authenticate separately to each IdP. Furthermore, the system should enhance trust relationships between different components and it should strengthen a privacy and security aspects.